\def\year{2022}\relax
\documentclass[letterpaper]{article} 
\usepackage{aaai22}   
\usepackage{times}  
\usepackage{helvet} 
\usepackage{courier}  
\usepackage[hyphens]{url}  
\usepackage{graphicx} 
\def\UrlFont{\rm}  
\usepackage{natbib}  
\usepackage{caption} 
\usepackage{enumerate}
\usepackage[show]{chato-notes}
\usepackage{booktabs}
\usepackage{listings}
\usepackage[utf8]{inputenc}
\usepackage{comment}
\usepackage{mathptmx}
\usepackage{latexsym,amsmath}
\usepackage{amsbsy}
\usepackage{amssymb}

\title{Echoes through Time: Evolution of the Italian COVID-19 Vaccination Debate}
\author {
    Giuseppe Crupi, Yelena Mejova, Michele Tizzani, Daniela Paolotti, Andr\'{e} Panisson\\
}
\affiliations {
    ISI Foundation, Turin, Italy\\
    gius.crupi@gmail.com,\{yelena.mejova,mt.tizzani,daniela.paolotti,andre.panisson\}@isi.it 
}

\begin{document}

\maketitle


\begin{abstract}

Italy was the first European country to be hit by COVID-19 in the early 2020, since then losing over 100,000 people to the disease. 
By the end of the vaccination campaign of 2021, 81\% of the public received at least one dose. 
These dramatic developments were accompanied by a rigorous discussion around vaccination, both about its urgency and its possible negative effects.
Twitter is one of the most popular social media platforms in the country, but pre-pandemic vaccination debate has been shown to be polarized and siloed into echo chambers.
It is thus imperative to understand the nature of this discourse, with a specific focus on the vaccination hesitant individuals, whose healthcare decisions may affect their communities and the country at large.
In this study we ask, how has the Italian discussion around vaccination changed during the COVID-19 pandemic, and have the unprecedented events of 2020-2021 been able to break the echo chamber around this topic?
We use a Twitter dataset spanning September 2019 - November 2021 to examine the state of polarization around vaccination.
We propose a hierarchical clustering approach to find the largest communities in the endorsement networks of different time periods, and manually illustrate that it produces communities of users sharing a stance.
Examining the structure of these networks, as well as textual content of their interactions, we find the stark division between supporters and hesitant individuals to continue throughout the vaccination campaign. 
However, we find an increasing commonality in the topical focus of the vaccine supporters and vaccine hesitant, pointing to a possible common set of facts the two sides may agree on.
Still, we discover a series of concerns voiced by the hesitant community, ranging from unfounded conspiracies (microchips in vaccines) to public health policy discussion (vaccine passport limitations).
We recommend an ongoing surveillance of this debate, especially to uncover concerns around vaccination before the public health decisions and official messaging are made public.

\end{abstract}

\section{Introduction}
\label{sec:intro}


The COVID-19 pandemic started in Whuan, China at the end of 2019 and in the course of a few months has spread globally, infecting hundreds of millions people around the world and killing more than five million since its beginning.
The unprecedented scale of the this pandemic has prompted the largest vaccination campaign of modern times: as of the end of 2021, 58.8\% of the world population has received at least one dose of a COVID-19 vaccine,
9.33 billion doses have been administered globally, and 30.42 million are now administered each day \cite{ourworldindata2022vaccinations}.
Whereas many parts of the world are yet to receive adequate number of vaccine doses, in the developed world, \emph{vaccine hesitancy} is becoming the major obstacle to full community coverage \cite{kottasova2021they}.
Latest metareviews of surveys show that vaccine acceptance rates range widely across Europe \cite{cascini2021attitudes}, some substantially below the percentage of the population that needs to be immunized to reach heard immunity (estimated at 82.5\% in May 2021 \cite{ke2021estimating}).
As the number of unvaccinated patients with COVID-19 rises in hospitals around the world, the European Commission President Von der Leyen has named the situation at the end of 2021 the `pandemic of the unvaccinated' \cite{herszenhorn2021von}.


Vaccination hesitancy has been an important public health issue even before COVID-19: for instance, it was included in the top 10 threats to global health in 2019 by the World Health Organization.\footnote{\url{https://www.who.int/emergencies/ten-threats-to-global-health-in-2019}}
Studies have found that vaccine hesitancy often occurs in a broader context of alternative health practices~\cite{Kalimeri2019}, science denial~\cite{browne2015going}, and belief in  conspiracy theories~\cite{jolley2014effects}.
On social media, it is driven by vocal influencers, who provide a wealth of content for the vaccine hesitant community to circulate~\cite{germani2021anti}.
Previous research has shown that social media sites, and Twitter in particular, are prone to siloing their users in ``echo chambers'' wherein users are mostly exposed to the opinions and information from like-minded individuals (including on the topic of vaccination~\cite{cossard2020falling,schmidt2018polarization}).
Worryingly, these echo chambers can support the propagation of potentially erroneous information \cite{chou2018addressing}.
In the light of the heated discussion around the COVID-19 vaccinations, social media platforms are beginning to issue bans on ``medical misinformation about currently administered vaccines that are approved and confirmed to be safe and effective'' 
(YouTube \cite{youtube2021misinfo}, Twitter \cite{twitter2021misinfo} and Facebook \cite{facebook2021misinfo}).
However, there are fears that such actions may stifle legitimate concerns over the safety and deployment of the COVID-19 vaccines, giving the censorship power to private companies~\cite{armitage2021online}. 
This battle over public opinion impacts the health-related decisions of countless individuals, potentially contributing to the rising death toll of the pandemic. 
\emph{Thus, we ask, have the events of 2020-2021 substantially transformed the vaccination debate? Have the unprecedented experiences and measures around a worldwide pandemic affected the echo chambers of vaccination hesitancy?}


In this work, we consider Italy as a case study -- the first European nation to detect COVID-19, institute lockdowns, and among the first ones to pay the highest human cost (mostly in the north of the country). 
On December 27, 2020 Italy launched its COVID-19 vaccination program, since then leading countries in Europe in terms of vaccine uptake~\cite{ourworldindata2022vaccinations}. 
Still, vaccine hesitancy is a threatening phenomenon, as the rate of fully vaccinated eligible individuals settled at 74\% at the end of 2021.
As many of its neighbors, by then the Italian government began requiring vaccination certificate (or ``green pass'') for most public activities, leading to political upheaval and social violence~\cite{local2021analysis}. 
In this study, we examine how the heated Twitter debate around vaccination in Italy has evolved during the roll out of the COVID-19 vaccination drive, and whether the traumatic events of the pandemic have unified the discussion around a common set of events.

Using a dataset spanning pre-pandemic vaccination debate, as well as the deployment of the nation-wide vaccination drive, we compare the state of the division among Italian speakers around the topic of vaccination.
We propose a flexible community detection approach to find a small number of the largest communities in the retweet network, and verify manually the opinion leaning of their members.
We then provide a temporal analysis of different aspects of the found communities to assess whether they have the characteristics of echo chambers: (i) retweet connectedness, (ii) attention via mentioning, (iii) and topical analysis of their tweets.

Despite the unprecedented events surrounding the COVID-19 pandemic, we find the echo chambers-like features to persist throughout the vaccination drive, with a negligible number of users changing their opinion over time.
We find that the vaccination hesitant community is not only vastly more vocal than the vaccine supporting one, it continued to grow throughout the later periods of study, indicating that new participants are joining the conversation every month.
Thus, the echo-chamber around vaccination hesitant not only persists, but grows substantially in size, suggesting that public health communication campaigns should not only focus on the main vaccination drive, but also on the latter periods when additional persuasion may be needed to achieve herd immunity.
Although we do find a shift in topical focus of both sides towards statistics around COVID-19 cases and vaccinations towards the end. 
This finding suggests that there may be a common ground of shared facts which may be used to bridge the gap between the communities. 
Interestingly, the vaccine hesitant community focuses on issues that prove to be contentious often months before the regulations become official, including vaccination of children and regulations around vaccination passports, which further suggests careful consideration of these early concerns may help in crafting public health messages.

We call upon the research and public health practitioners to continue the monitoring of those voicing concerns around vaccination in order to better understand the new concerns and perspectives, as well as the way social media echo chambers may affect the nature of this conversation.

\section{Related Work}
\label{sec:rel_work}



\textbf{Vaccine Hesitancy.} Prior to the protests around the COVID-19 vaccination campaigns, the rise in vaccination hesitancy has been largely associated with the reluctance of parents to vaccinate their children \cite{kieslich2018addressing} and the attitude of adults with respect to seasonal flu vaccination.
Compared to North America and North European countries, confidence in the safety of vaccination is especially low in Western and Eastern Europe, with 59\% and 40\% respectively agreeing that vaccines are safe~\cite{wellcome2018attitudes}.
It has also been found that parents who deviate from recommended childhood vaccination schedules are more likely to focus on potential serious adverse reactions to vaccination~\cite{betti2021detecting} and display lower trust in science~\cite{wellcome2018attitudes}.
Vaccine hesitancy is not only related to reluctance to engage with scientific evidence~\cite{browne2015going}, but it has also been linked to the emphasis on religious identity~\cite{kata2010postmodern},
alternative or holistic health methods~\cite{Kalimeri2019,betti2021detecting}, and certain political attitudes~\cite{yaqub2014attitudes,browne2015going}.
However, public health communication campaigns may be producing an effect. 
The 2020 Vaccine Confidence Project report stated that, compared to 2018, a growing majority of the EU and UK public agreed that vaccines are important, effective, safe and compatible with their religion~\cite{de2020state}. 


It is important to understand whether these attitudes persist in the midst of the COVID-19 pandemic. A systematic study in December 2020 has found that Italy had some of the lowest vaccine acceptance rates (53.7\%), on par with Russia (54.9\%), Poland (56.3\%), and US (56.9\%)~\cite{sallam2021covid}.
However, there is some disagreement about the temporal variability in the perception of Italians of vaccination during the pandemic.
\citet{palamenghi2020mistrust} find that the trust in research and in vaccines decreased in the Italian public between the beginning of the pandemic and the end of the first lockdown, whereas \citet{caserotti2021associations} find that more respondents were more willing to get vaccinated for COVID-19, regardless of their beliefs about vaccines, in around the same time frame (though, both estimate that around 59\% of Italians were willing to vaccinate against the disease).
Hesitancy remained at 31\% in the north of Italy in January 2021, which suffered the worst of the pandemic \cite{reno2021enhancing}, and even those discharged from hospitals recovering from COVID-19 were hesitant or undecided towards vaccines at 59\% \cite{gerussi2021vaccine}.
However, these attitudes may vary greatly within the population. 
For instance, as vaccines are being approved for younger children, surveys show that 82\% of Italian parents were willing to vaccinate their child of 12-18 years \cite{bianco2021parental}.


In the context of vaccination hesitancy, and especially during the pandemic, social media has been deemed as a major source of community organization and information exchange. 
As COVID-19 struck Italy, its residents turned to the internet to learn about various conspiracies, such as YouTube channels and Google: \citet{rovetta2021impact} finds that the interest in vaccine side effects even exceeded interest in pollution and climate change.
In fact, some of the interest in ``fake news'' topics was spurred by the local mainstream media \cite{rovetta2021influence}.
Similarly, a study in UK has found that a combination of social media dependence and high levels of conspiracy mentality are most likely to be associated with online discouragement of vaccination \cite{chadwick2021online}.
Moreover, a complex interplay between lack of trust, conspiracy beliefs, and social media use has found to be a strong predictor of COVID-19 vaccine hesitancy in the UK ~\cite{vaccines9060593}.
In Middle East and Gulf countries, those relying on social media platforms as the main source of information were likely to score higher on the Vaccine Conspiracy Belief Scale (VCBS)~\cite{sallam2021high}. 
Thus, research is ongoing to monitor major social media sites in order to capture community interactions of users interested in vaccination, and to reveal the potential impact of misinformation~\cite{pierri2021vaccinitaly}. 
Unlike these efforts, our data collection begins before the onset of COVID-19, potentially capturing the thematic and behavioral changes of the anti-vaccination movement during this unique time.

\textbf{Echo Chambers}. Beyond misinformation, the very structure of discourse on social media, and especially Twitter, may foster division. 
First highlighted in the US political sphere \cite{conover2011political}, divisions both in content and network structure have been revealed in numerous discussions around controversial topics~\cite{garimella2018political,del2016echo,garrett2009echo}. 
These discussions are often characterized by homogeneity in the way the information spreads within the group, supporting distinct and sometimes radical opinion formation~\cite{del2016spreading,cota2019quantifying,garimella2017effect}.
A temporal study of the attention dynamics also found that an increase in attention to the topic increases the polarization between those with different opinion~\cite{garimella2017effect}. 
If this trend can be confirmed in the case of vaccination debate, such increase in division may damage the public health efforts when they are most urgently needed.
Recently, \citet{cossard2020falling} showed that the pre-pandemic Italian vaccination debate was already highly polarized.
In this work, we inspect the dynamical nature of the potential polarization around vaccination in the duration of one of the largest vaccination campaigns in Europe.

The detection of differing opinions and communities around these opinions is an active research direction. 
\citet{li2019opinion} propose an opinion community detection method by considering the content similarity, the time similarity and the topology structure of users, which they then applied to a discussion forum.
\citet{sinha2020hierarchical} use hierarchical clustering algorithm for the characterization of users according to their temporal activity, achieving topical heterogeneity in the clusters.
For Twitter specifically, \citet{sanchez2016twitter} propose to use the modularity-based Louvain algorithm to detect users with similar political stance.
Since the greedy optimization employed by Louvain often produces many small communities in order to maximize the modularity of the outcome, researchers interested in outlining two major sides of a debate proposed a bipartitioning approach.
\citet{garimella2018quantifying} apply a graph partitioning algorithm METIS~\cite{karypis1998fast} to retweet networks of discussions around potentially controversial topics.
They then propose a Random Walk Controversy score (RWC) that measures how connected the two partitions are by performing personalized random walks from authoritative nodes of each side.
It has been used to quantify the controversy around topics spanning gun control~\cite{ozer2019measuring}, Supreme Court nominations~\cite{darwish2019quantifying}, as well as the vaccine debate~\cite{cossard2020falling}.
In this work, we do not assume that there are only 2 sides of the debate, and propose an alternative opinion community detection approach based on hierarchical clustering, in order to illustrate the changing dynamics of the discussion over time.

\section{Methods and Results}

\subsection{Data Collection and Description}
\label{sec:method}

To collect our data, we query the public Twitter Streaming API, filtering with Italian language and the following keywords: \textit{vacc, vaccinale, vaccinali, vaccinano, vaccinarci, vaccinare, vaccinarsi, vaccinate, vaccinati, vaccinato, vaccinaz, vaccinazione, vaccinazioni, vaccines, vaccini, vaccinista, vaccinisti, vaccino, antivaccinisti, freevax, iovaccino, nonvaccinato, novax, obbligovaccinale, provax, ridacciilvaccino} (which cover the keyword ``vaccine'' in different linguistic forms and related terms), as well as the same terms in English. 
The resulting collection spans from September 5th, 2019 to November 7th, 2021, encompassing a totality of 795 days and including 16,223,749 tweets from 665,882 unique users.
Note that the API provides only publicly shared tweets, i.e. no tweets sent as direct messages or posted by private accounts are available.
Also, we do not use a geographic constraint, thus the users may reside anywhere on earth, but due to the limited use of the Italian language outside of Italy\footnote{\url{https://www.britannica.com/topic/Italian-language}} (compared to English, Spanish, and Portuguese), we assume that the majority of the captured users are indeed residents of Italy.


Notably, this data set captures the pre-pandemic vaccination debate, the reactions around the various events surrounding COVID-19 as well as discussions around the development and deployment of COVID-19 vaccines. 
We therefore segment the data timeline into six periods that encompass various historical moments around vaccination in Italy: 

\begin{enumerate}[i]
\item \textit{pre-Covid}: before the advent of the pandemic (2019/9/5 - 2019/12/31); 
\item \textit{early-Covid}: when novel coronavirus was being tracked in Wuhan, China (2020/1/1 - 2020/3/8); 
\item \textit{pre-vaccine}: between the first Italian lockdown and the advent of the vaccines (2020/3/9 - 2020/10/31); 
\item \textit{early-vaccine}: during the first months of the COVID-19 vaccination campaign (2020/11/1 - 2021/4/16); 
\item \textit{vaccine-drive}: the main vaccination drive (2021/4/17 - 2021/7/31);
\item \textit{late-vaccine}: once a significant fraction of the Italian population was fully vaccinated (2021/8/1 - 2021/11/7). 
\end{enumerate}


\begin{figure} 
\centering
\includegraphics[width=\columnwidth]{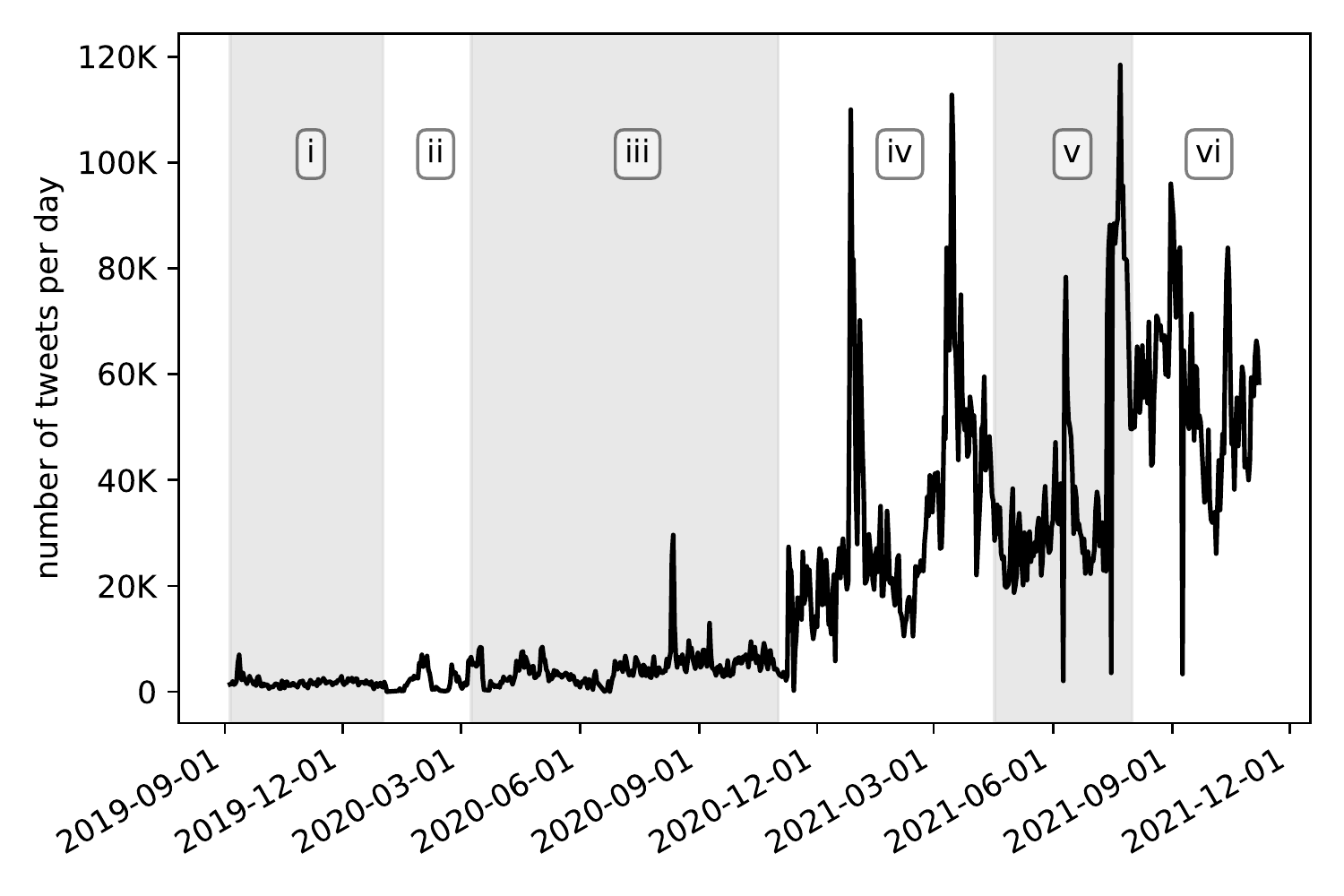}
\caption{Dataset volume by day. Vertical lines delineate time periods.}
\label{fig:volume}
\end{figure}

Figure \ref{fig:volume} shows the daily volume of tweets that have been collected. In the pre-pandemic period, the volume remained at an average of 1,757 tweets per day. Unfortunately, during this period, we experienced some technical difficulties which led to a total of 22 days with missing data.
The volume of daily tweets about vaccination starts rising at the end of summer 2020 and reaches a first small peak at the beginning of November 2020, when industrialized countries (including Italy) were approaching the beginning of the vaccination campaign.
The first major peak was observed in correspondence of the Italian Vaccine Day (2020/12/27). 
Since then, the debate becomes even more heathed and picks up volume throughout the rest of the data collection window. 
The second peak is in mid March, corresponding to the temporary suspension of the AstraZeneca vaccine by the Italian Medicines Agency (AIFA) \footnote{\url{https://www.aifa.gov.it/-/aifa-sospensione-precauzionale-del-vaccino-astrazeneca}}.
Other smaller peaks occur on days devoted to open COVID-19 vaccination and flu vaccination campaign launch in different regions \footnote{\url{https://en.wikipedia.org/wiki/COVID-19_vaccination_in_Italy}}. 
Finally, the last of the higher peaks is due to the statement made at the press conference by the Italian Prime Minister Mario Draghi, when he mentioned that the appeal not to be vaccinated is equivalent to an appeal to die \footnote{\url{https://www.ansa.it/english/newswire/english_service/2021/07/23/appeal-against-vaccination-an-appeal-to-die-says-draghi-10_92332c59-2145-44e8-b58a-639678625741.html}}.

\subsection{Network Construction}

Retweet networks have been used extensively to study controversial topics on Twitter \cite{stewart2018examining} as retweets are usually considered a form of endorsement among users. 
A retweet network is a weighted directed network where nodes represent users and the weight of an edge from node \textit{u} to node \textit{v} represents the number of times that user \textit{u} retweeted user \textit{v}. 
In order to study the unfolding of the debate around vaccines, we build six retweet networks, one for each of the above mentioned time periods. 
Following previous literature, we exclude all edges with a weight equal to one to reduce the noise in the data \cite{garimella2018quantifying}. 
For each network, we extract the largest Weakly Connected Component (WCC). 
The sizes of the WCCs are reported in Table \ref{tab:networks_size}. 
In total, the networks capture 92,840 unique users, and 6,300,540 tweets.


\begin{table}
    \centering
\begin{tabular}{rl|rrr}
     \toprule
     && \textbf{$\mid V \mid$} & \textbf{$\mid E \mid$}& \textit{\# tweets}\\
     \midrule
     i. & pre-Covid & 5,528 & 18,409 & 92,560\\
     ii. & early-Covid & 4,247 & 9,054 & 29,237\\
     iii. & pre-vaccine & 18,967 & 80,234 & 348,887\\
     iv. & early-vaccine & 59,398 & 410,515 & 1,911,784\\
     v. & vaccine-drive & 43,325 & 318,284 & 1,596,052\\
     vi. & late-vaccine & 44,840 & 451,118 & 2,322,020\\
     \bottomrule
\end{tabular}\\
    \caption{Size of the WCC of retweet network by time period.}
\label{tab:networks_size}
\end{table}


\subsection{Community Detection}

Since retweets are usually interpreted as an endorsement of the opinion expressed in the retweeted message, they can be used to identify groups of Twitter users which take a particular stance in a public debate \cite{garimella2018quantifying}. We thus investigate the topological structure of the retweet networks in order to identify the attitudes and opinions users in the vaccination debate. 
To this goal, we use community detection to find groups of users who share an opinion. 
We use \textit{Paris}, an agglomerative hierarchical clustering algorithm \cite{bonald2018hierarchical}, as it provides flexibility in the selection of the number of communities in each network. 
In particular, the agglomerative hierarchical clustering algorithm provides us with a dendrogram of successive partitions of the network, with each level being associated with a specific number of communities.
For each partition we can compute the modularity score, and select the optimal partition as the one with  the first modularity increase by at least 10\%.

\begin{figure} 
\centering
\includegraphics[width=\columnwidth]{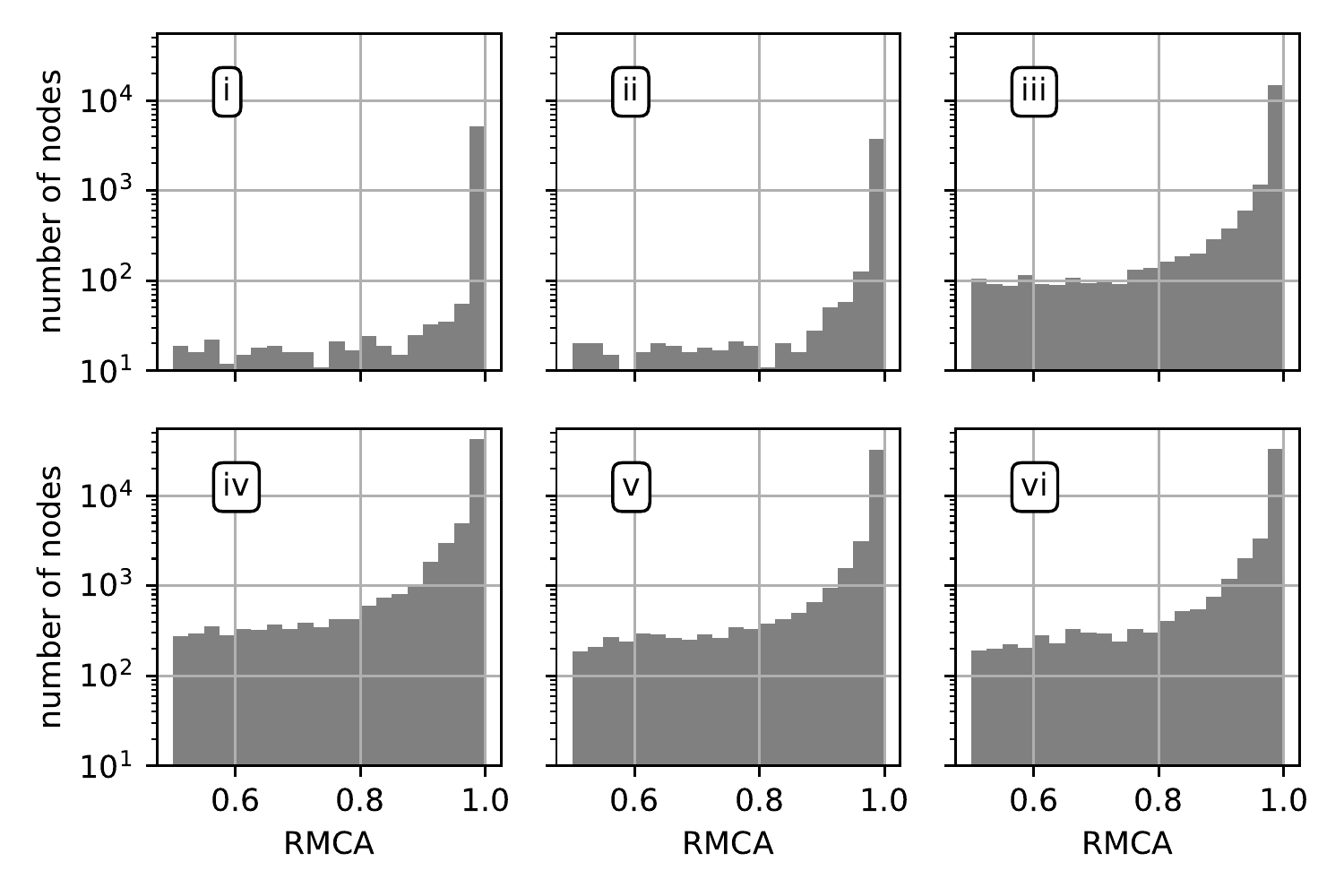}
\caption{RMCA distributions for each time period.}
\label{fig:RMC}
\end{figure}

In order to detect a statistically robust partition of the retweet networks, we use the following approach:
(i) we sample the given network by removing half of its edges; (ii) we run the Paris algorithm on the Weakly Connected Component of the sampled graph; (iii) we select the communities using the modularity criterion above. These three steps are then repeated 100 times with independent samples. 
In order to have a reasonable confidence that a sampled node should be assigned to a given community, we use the fraction of time the node has been assigned to the community to which it was assigned most frequently (we dub this measure as the Rate of the Most Common Assignment (RMCA)). 
The distributions of RMCA scores are reported in Figure \ref{fig:RMC}. We find that the communities found by the Paris algorithm are robust and that the majority of the users are assigned to the very same community most of the time. 
Based on this observation, we set the RMCA threshold for assignment at 0.9; then we name as \textit{assigned} the users/nodes which have an RMCA above the threshold and \textit{unassigned} the users/node with an RMCA below the threshold. With this choice we are able to assign 87.6 – 94.8\% of the nodes in each time window. 
We note that the fraction of nodes which is included in the sample less than 30 times out of 100 is 1.4 – 4.4\% for each time period, resulting in a sufficient information for the assignment for the vast majority of the nodes.

\begin{figure} 
\centering
\includegraphics[width=\columnwidth]{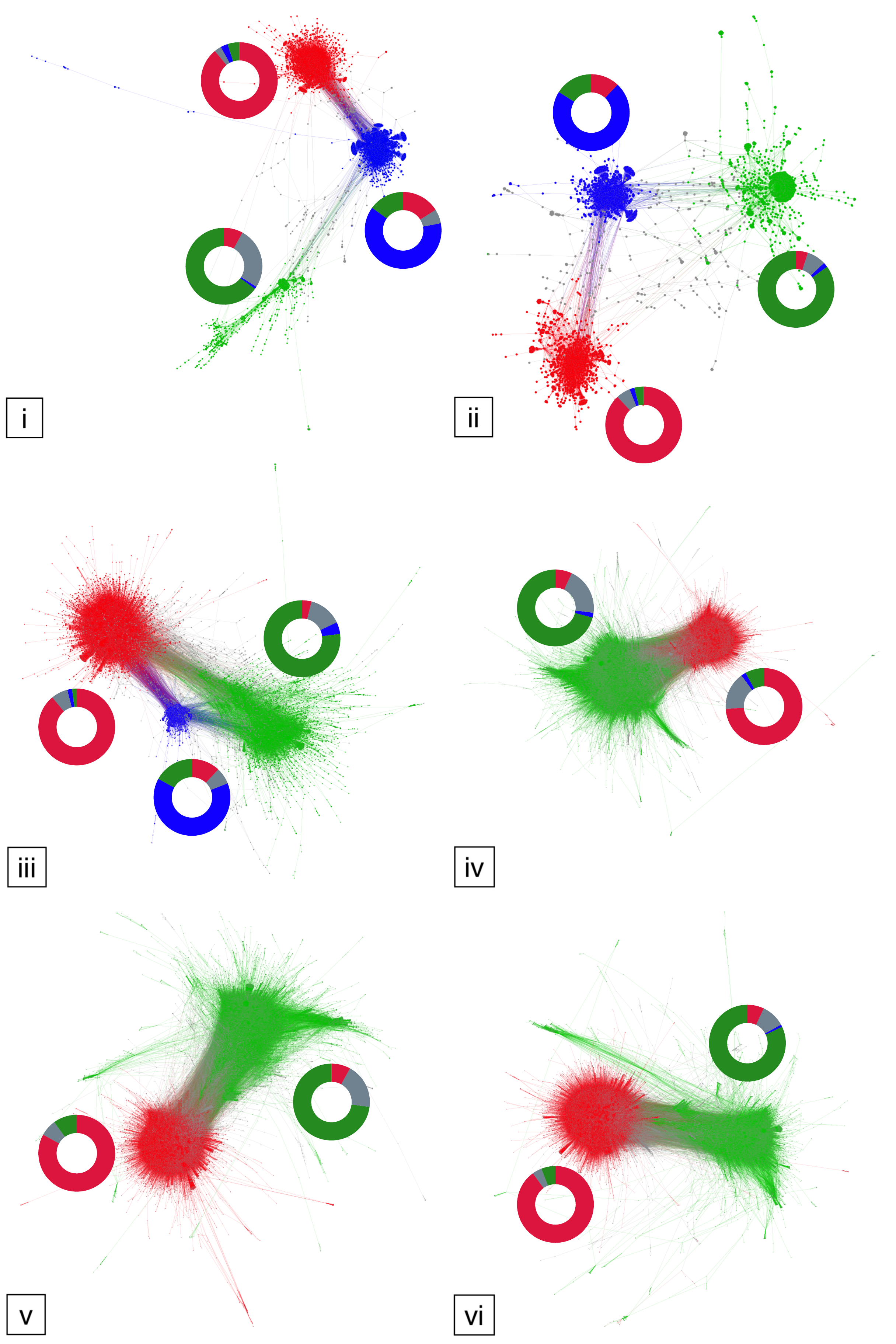}
\caption{Retweet networks for each time period, colored by community (red: hesitant, green: supportive, blue: pets, grey: other).}
\label{fig:networks}
\end{figure}

\begin{figure} 
\centering
\includegraphics[width=\columnwidth]{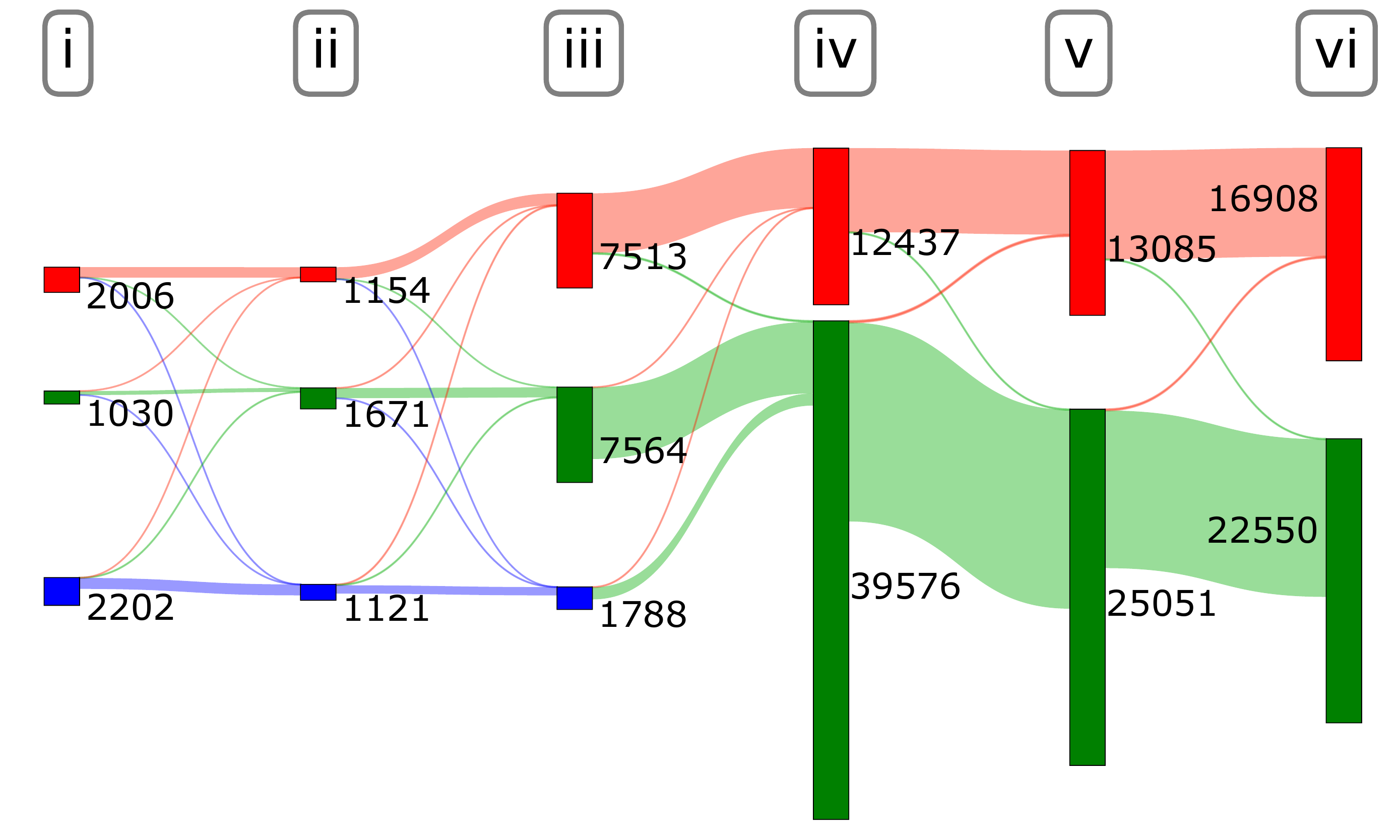}
\caption{Flow diagram for community evolution. Vertical layers represent time segments, while colored boxes represent polarized communities (red: hesitant, green: supportive, blue: pets). The numbers indicate the size of the communities.}
\label{fig:flow_diagram}
\end{figure}

In Figure \ref{fig:networks} we show the six retweet networks, plotted using Gephi \cite{bastian2009gephi} with the force-directed ForceAtlas2 algorithm \cite{jacomy2014forceatlas2} applied for the layout, and colored by community (characterized in the next section).
We find that the debate has well-defined communities, and is divided into 3 clusters for the first three time segments, and 2 in the last.
The structure of these networks, especially in the early periods, exemplifies the echo chamber effect wherein users retweet mostly others from the same community, and infrequently those from the others. 
The size of the communities can be seen in Figure \ref{fig:flow_diagram}: the largest engagement occurs at the early stages of the vaccine roll-out in period \emph{iv}.

\subsection{Opinion Annotation}

To understand the content posted by the users in each community in each time period's network, we manually label a sample of users and their tweets from each.
For each community in a given time segment, we sample 100 users and up to 5 tweets per user in that specific time window. There is no users' overlap across communities in different time windows.
This procedure resulted in a sample of 1500 users. 
Four Italian annotators labeled the users as ``vaccination supporter'', ``vaccination hesitant'', ``other'' (used for users not having a clear stance), and ``pets'' category dedicated to those referring to pets and animals. 
Note, that we use the broader definition of ``vaccination hesitant'', rather than ``anti-vaccination'' stance, as we would like to capture both those who are expressing hesitation around vaccines, as well as those opposed to vaccination in general, since it is difficult to distinguish these stances from a small sample of tweets (see Discussion for more on the spectrum of opinion).
The task proved to be moderately easy, with an overlap of 85\% in labels, or a Cohen's Kappa of 0.72, which is relatively high, considering a 4-category task.
We note that all disagreements were with the ``other'' class, and not between the supporting and hesitant sides, illustrating that tweets usually provide enough information to distinguish the user's stance on this topic.

The circles in Figure \ref{fig:networks} show the share of each category in the annotated communities, in each time period.
We find that the membership of each community is composed of a majority of one class, with majority classes ranging from 63\% - 90\%.
Thus, we use these annotations to propagate the opinion label throughout each community, with each annotated sample providing an estimate of the accuracy of this approach (which is quite high especially in the last few time periods). 

Also in Figure \ref{fig:networks}, we find that in the first few periods, the pet community was a prominent bridge between the supporting and hesitant communities.
However, as the focus and volume of the conversation grows around COVID-19, the pet group is not significant enough to be identified by our algorithm.
Although the supporting community grows over pre- and early-vaccine time periods, 
in the last two time windows the hesitant community gains importance (in the last time window, almost 40\% of the users are hesitant). 

Both sides are organized around their several authoritative and/or popular users. The most retweeted user of the supporters community is always Roberto Burioni (@RobertoBurioni), an Italian virologist, physician, university professor and pro-vaccination activist. Among recurrent hubs of the supporters side we also find doctors (@Cartabellotta) and official accounts of Health Institutions (@MinisteroSalute).
On the vaccine hesitant side we find anonymous accounts claiming no relevant medical qualifications, such as those advocating against children vaccination (@MinervaMcGrani1), accounts sharing fierce government opposition content (@intuslegens and @RadioSavana), as well as suspended accounts (@EuroMasochismo). 
Despite being anonymous, these accounts have tens of thousands followers, and their tweets often reach up to thousand retweets.




\subsection{Quantifying the Echo Chamber over Time}

As the network visualizations illustrate, the Italian vaccination debate is a polarizing subject. Below, we quantify the extent of this polarization between the communities, as well as over time. 

\begin{figure} 
\centering
\includegraphics[width=0.9\columnwidth]{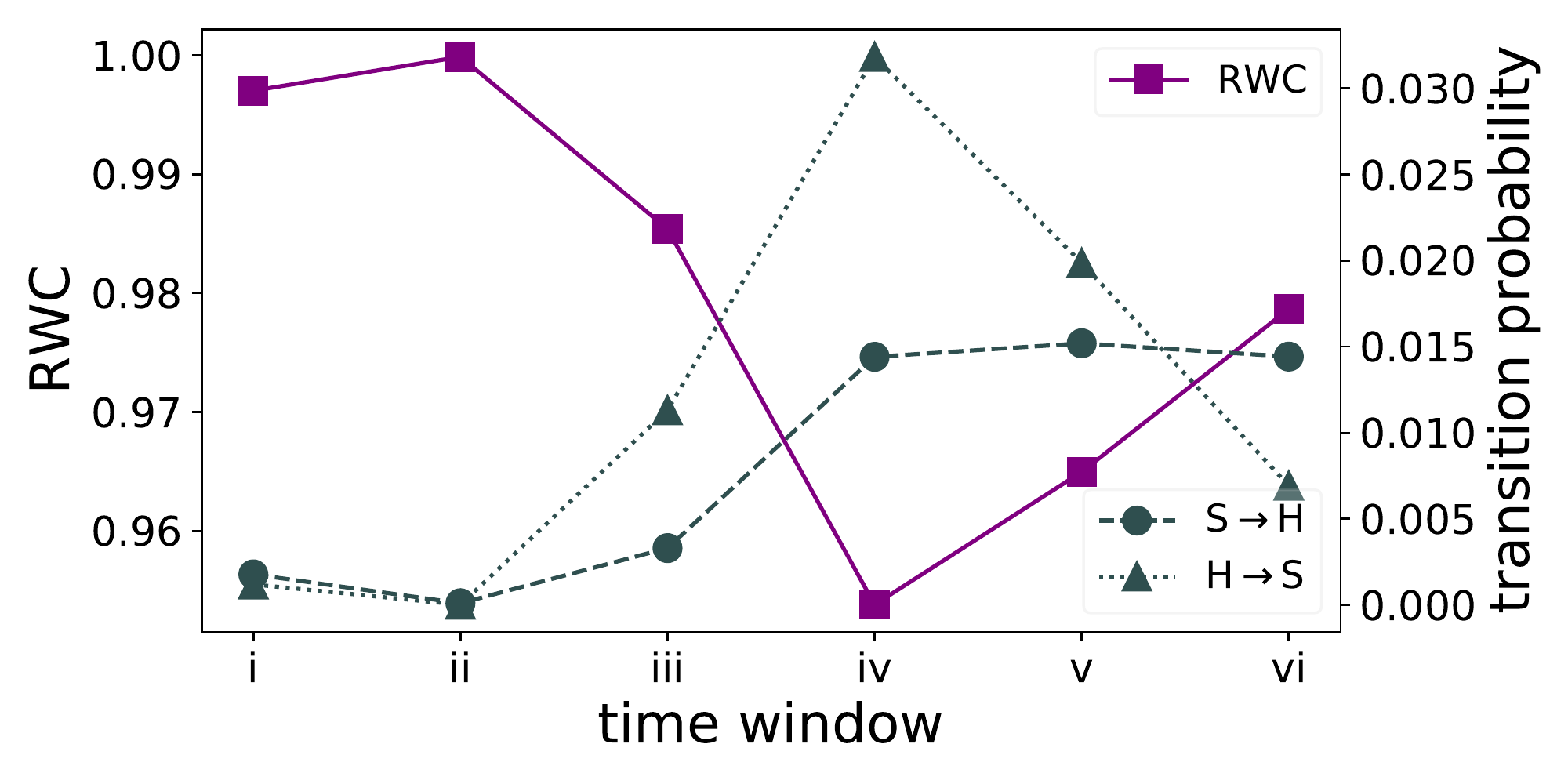}
\caption{Random Walk Controversy score (left y axis) and probabilities of starting on one side, given walk ends on another side (right y axis), for the six time periods.}
\label{fig:rwc}
\end{figure}

We utilize the Random Walk Controversy score (RWC) proposed by \citet{garimella2018quantifying} in order to quantify the extent to which the two sides of the argument are disconnected. 
Intuitively, it measures ``how likely a random user on either side is to be exposed to authoritative content from the opposing side''.
To prepare the graphs, we first remove the users identified as interested in pets, as their content is not relevant to the focus of this study.
Note that we keep the users who were not assigned to a community (those shown in grey in Figure \ref{fig:networks}), as they often appear between the two main clusters, and provide a bridge between the users assigned a community.
The RWC score (with parameter $k=100$) for each time segment is shown in Figure \ref{fig:rwc}, as well as the probabilities that one side retweets the other (the probabilities that a side retweets its own side is almost always close to 100\%, and for clarity they are not plotted).
We find that the two sides are always very separated, with RWC ranging between 0.95 and 0.99. 
The controversy score is the lowest during the early-vaccine period (\emph{iv}), when the probability is highest that one has started the random walk in the hesitant community, given one reached the supporting one.
It is possible that during the roll out of the campaign, informational tweets were more likely to be retweeted by both sides, including the hesitant one (we confirm this during content analysis).
The controversy score goes back up again during the last two periods, suggesting return to the lack of interaction between the two sides.

\begin{figure} 
\centering
\includegraphics[width=0.8\columnwidth]{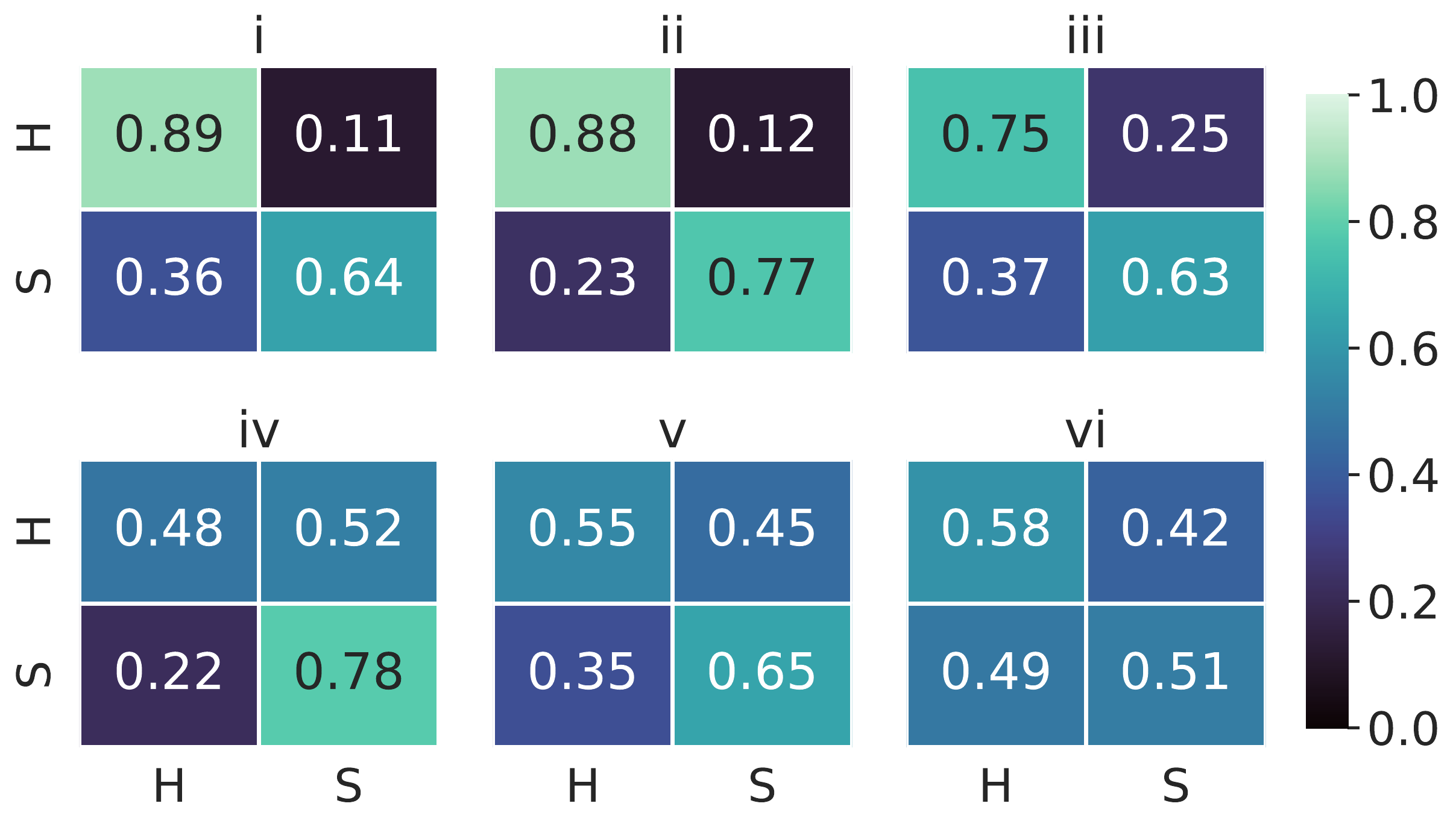}
\caption{Share of mentions of users in each time period: out of all mentions by row $x$, how many are from column $y$ (S - supporters, H - hesitant).}
\label{fig:mentions}
\end{figure}


Figure \ref{fig:flow_diagram} shows a Sankey diagram that illustrates the overlap of membership in the communities between the time periods. 
The number of users in each community is indicated by numbers.
We find a stable, rapidly increasing membership on both supporting and hesitant sides, with few users changing affiliation: fraction of users switching side is 0.7 – 4.4\% for every time window.
Further, a substantial proportion of users continue to be active from one period to another, indicating a continuous attention on the matter.
This is true not only during the COVID-19 period, but the vaccine hesitant communities in the pre-pandemic periods retain their stance during the pandemic, suggesting a consistency of opinion over time, despite the new developments introduced by COVID-19.
The average span of engagement (difference between first and last post in the dataset) of a supporting user is 415 days (median 399) and of hesitant one 360 days (median 332), indicating attention span of around a year.
The reader should keep in mind that these users are more active due to construction of the retweet networks, as the average span of engagement for any user in the dataset is 118 days (median 0, as most post only once). 
Still, we find that, in these communities of active participants, there is a marked continuity of membership: out of 2,335 vaccine hesitant users in the first two periods, 1,379 (59\%) also engage in the hesitant group in the last two periods (and only 44 (1.9\%) end up in the supporting group).

Alternatively, we may consider the extent to which users from one side mention the other, which would suggest some attention across the opinion divide.
Figure \ref{fig:mentions} shows the proportion of mentions that come from the opposite side from the supporter and hesitant groups.
We find the most divided attention in the first two periods, where hesitant users tend to not mention any supportive users.
However, this attention trend reverses over time, with the last two periods having a more even distribution of mentions -- by both sides, of the other side. 
Thus, in the dimension of mentioning the other side, the COVID-19 vaccination campaign has reduced the echo chamber effect in the debate.

In summary, we find that, in terms of stance and retweeting, the two sides of the debate remain separate in the duration of the study. 
However, the fact that the mentions across the aisle increase suggests a shift in attention to a common set of actors, and perhaps topics.
In the next section, we compare the content of the posts and provide a qualitative evidence of different foci of the two sides of the debate.

\subsection{Topic Variation over Time}

\begin{table}[t]
    \centering
\begin{tabular}{rl|rrrr}
     \toprule
     &&  \textbf{$\mid H_{usr} \mid$} & \textbf{$\mid S_{usr} \mid$} & \textbf{$\mid H_{day} \mid$} & \textbf{$\mid S_{day} \mid$}\\
     \midrule
     i. & pre-Covid & 41.9 & 8.7 & 0.35 & 0.07\\
     ii. & early-Covid & 24.0 & 7.0 & 0.40 & 0.12 \\
     iii. & pre-vaccine & 61.8 & 18.8 & 0.27 & 0.08\\
     iv. & early-vaccine & 137.4 & 48.7 & 0.83 & 0.30 \\
     v. & vaccine-drive & 153.2 & 37.4 & 1.45 & 0.35\\
     vi. & late-vaccine & 191.3 & 43.2 & 1.95 & 0.44\\
     \bottomrule
\end{tabular}\\
    \caption{Average posting frequency in each time period by vaccine hesitant (H) and supportive (S) users. Left two columns (usr): average number of posts per user in time window, right two columns (day): same, per day.}
\label{tab:tweet_per_user}
\end{table}

As shown in previous literature \cite{cossard2020falling}, the users labeled as vaccine hesitant post more content than others, being more ``vocal''. 
Table \ref{tab:tweet_per_user} shows the average posting rate per user in each time window (left two columns), as well as normalized by day, which controls for the size of the time window (right two columns).
The disparity between the posting rate increases throughout the study duration: hesitant users post on average 0.35 tweets per day in period \emph{i}, and 1.95 tweets in period \emph{vi}, that is over 5 times more.
We ask, how different is the content produced by these two sets of users? That is, does the echo chamber reflected in the topics discussed by each side?


We use topic analysis framework to summarize the main topics discussed in each time window.
Although there are many approaches to temporal and hierarchical topic modeling~\cite{blei2006dynamic,panisson2014mining,gobbo2019topic}, the approach used here is similar to the approach  described in \citet{gozzi2020collective}.

We begin the content analysis by pre-processing the text of the tweets.
We remove urls, hashtag and mention signs, numbers, stopwords (both Italian and English stopwords), remove the query words and other common Italian words.
In addition to tokens composed of single words, we extract phrases using a PMI-like scoring as described in~\citet{mikolov2013distributed}.
We remove all tokens and phrases appearing in fewer than 10 tweets and those appearing in more than half of the tweets. 
We run this pre-processing step independently for each time window:
we extract the dataset $\mathcal{D}$ of tweets with the following numbers of tweets:
(i) 204k, (ii) 126k, (iii) 1M, (iv) 5.1M, (v) 4.1M, (vi) 5.5M, totalling in about 16M tweets.
The pre-processing step results in a vocabulary $\mathcal{V}$ in each time window with the following number of tokens and phrases:
(i) 7,458, (ii) 6,171, (iii) 32,842, (iv) 95,428, (v) 85,922, (vi) 100,285.




Each document is then represented as a vector of term counts, in a \textit{bag-of-words} approach. We apply TF-IDF normalization and, in each time interval, we extract a total of $K = 20$ topics through this NMF optimization: $\min\limits_{W,H} {\Vert \textbf{X} - \textbf{W} \textbf{H} \Vert}^2_F$,
where $\Vert  \Vert ^2_F$ is the Frobenius norm and $\textbf{X} \in \mathbb{R}^{|\mathcal{D}|\times |\mathcal{V}|}$ is the matrix resulting form TF-IDF normalization,
subject to the constraint that
the values in $\textbf{W} \in \mathbb{R}^{|\mathcal{D}|\times K}$ and $\textbf{H} \in \mathbb{R}^{K\times |\mathcal{V}|}$ must be nonnegative. 
The nonnegative factorization is achieved using the hierarchical alternating least squares algorithm described in \citet{cichocki2009fast}.
We apply this method to 
each time period separately.

The topic extraction in the interval-specific dataset $\mathcal{D}$ produces a set of general topics, discussed by users that are in both groups of vaccine hesitant and vaccine supporters.
For each of these topics, we calculate the share of activity from vaccine hesitant users. For this, we select from $\mathbf{W}$ only the rows corresponding to each of these two groups:
if we consider $\mathcal{H} \subset \mathcal{D}$ as the tweets from vaccine hesitant users and $\mathcal{S} \subset \mathcal{D}$ as the tweets from vaccine supporters, 
%
then we calculate the share of vaccine hesitant users $\varrho_k$ for a given topic $k$ as follows:
%
\begin{equation}
\setlength\abovedisplayskip{0pt}
\setlength\belowdisplayskip{5pt}
\varrho_k = 
\frac{
|\mathcal{S}| \cdot \sum_{i \in \mathcal{H}}{w_{ik}}
}
{|\mathcal{H}| \cdot \sum_{i \in \mathcal{S}}{w_{ik}} + 
|\mathcal{S}| \cdot \sum_{i \in \mathcal{H}}{w_{ik}}
}
\label{share}
\end{equation}
where $w_{ik}$ is the $i,k$ value in $\textbf{W}$.
The coefficients $|\mathcal{H}|$ and $|\mathcal{S}|$ are necessary to normalize the share such that if the probability of a vaccine hesitant user posting in a topic is the same as a vaccine supporter, then the share is equal to 0.5 independently of the proportion of users in each group.

In a further step, to analyse the topics discussed by each specific group, we select only the tweets from each group of hesitant and supporter users $\mathcal{H}, \mathcal{S}$ in new matrices $\widetilde{\textbf{X}}$ and extract also the 20 most prominent topics for each of them. 
The matrix $\widetilde{\textbf{H}}$ extracted via NMF and representing the group-specific topics is then used as a transformation basis for the whole dataset $\mathcal{D}$, i.e. with the whole matrix $\textbf{X}$ we fix $\textbf{H}$ to $\widetilde{\textbf{H}}$ and calculate a new $\widetilde{\textbf{W}}$ according to
the NMF optimization.
Finally, we calculate the share of vaccine hesitant users for each group-specific topic using Eq.~\ref{share} where $w_{ik}$ is the $i,k$ value in $\widetilde{\textbf{W}}$.

\begin{figure} 
\centering
\includegraphics[width=\columnwidth]{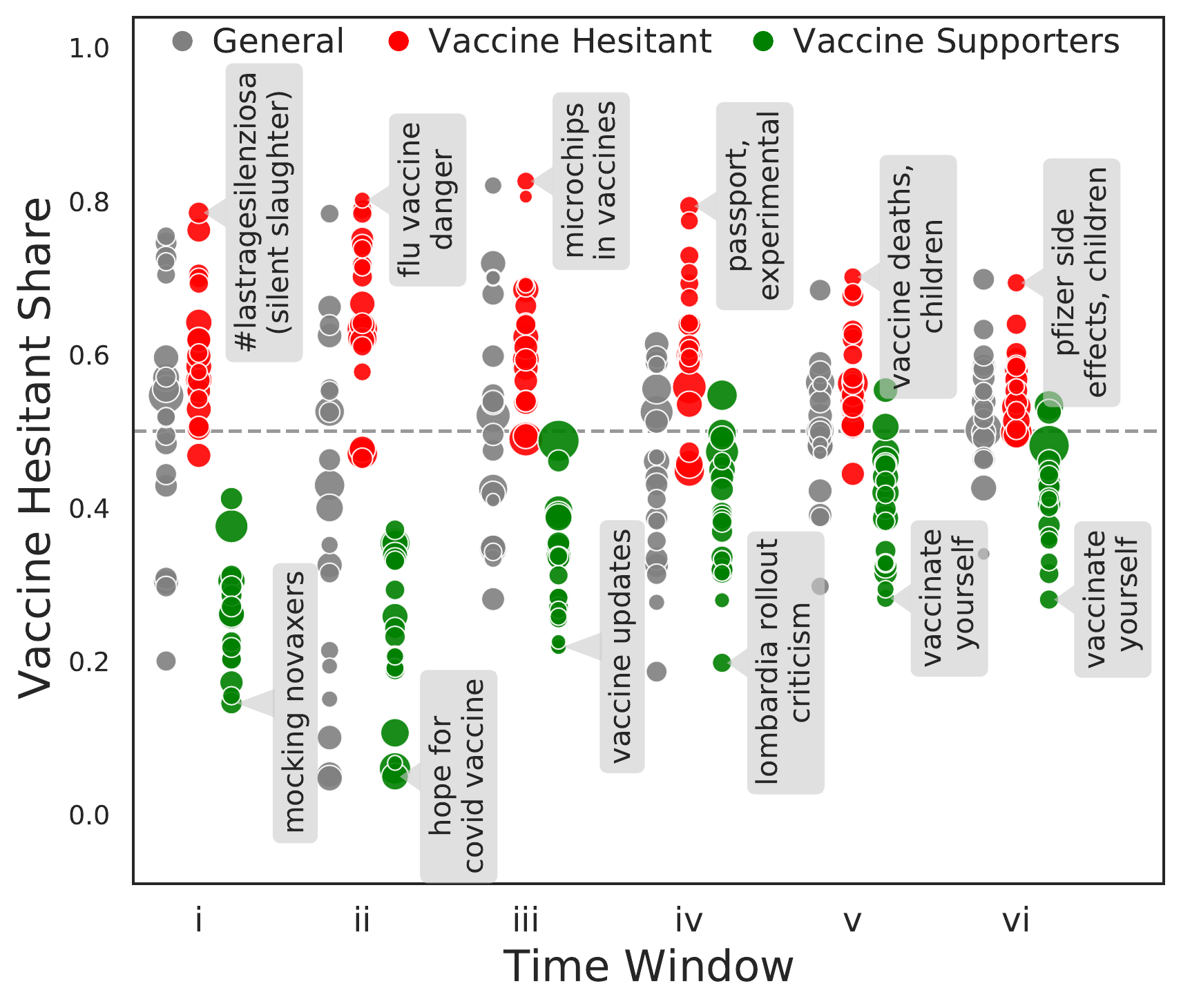}
\caption{Distribution of side share (proportion of tweets from vaccine hesitant labeled users) for the 20 most prominent topics for each time period, modeled using all users (grey), hesitant (red), or supportive (green), with size of point proportional to the importance of the topic.}
\label{fig:topic_side_share}
\end{figure}

In Figure \ref{fig:topic_side_share} we show the results of this interval-specific, group-specific topic modeling approach. Each dot is an extracted topic, positioned in their time window of reference. On the y axis we show the vaccine hesitant share of the topic $\varrho_k$, calculated according to Eq.~\ref{share}. The color shows the group of users from which the topics are extracted: the grey dots show the 20 most prominent topics extracted when using the full matrix $\textbf{X}$, the red dots are extracted using a matrix $\widetilde{\textbf{X}}$ with tweets from vaccine hesitant users, and the green dots are extracted using a matrix $\widetilde{\textbf{X}}$ with tweets from supportive users. The size of the dots represent their importance relative to the other topics in the same group, i.e., the relative importance of a topic $k$ is calculated as $ \frac{\sum_{i \in \mathcal{D}}{w_{ik}}}{\sum_{l}\sum_{i \in \mathcal{D}}{w_{il}}} $, where $w_{ik}, w_{il}$ are values of $\textbf{W}$ in the case of general topics, and are values in $\widetilde{\textbf{W}}$ in the case of the group specific topics.
Finally, we add an annotation to the topics with the highest and lowest vaccine hesitant shares in each time window, only for the topics extracted from hesitant and from supporting users.


As expected, the topics extracted from one side (say, supporters) are indeed more prominent in the content of the supporters (the green dots are closer to 0). 
This separation in topic discussion is especially evident in period \emph{ii}, when the side share is rarely close to 50/50. 
As the time goes on, the polarization around the topics declines, and more topics appear in tweets from both sides.
The common topics in the last two periods include statistics about COVID-19 cases and vaccinations, as well as discussions around vaccinating young people and vaccination passport rules.

The figure also contains summaries of topics with most leaning to one or the other side, obtained via manual examination of the topics and associated tweets.
For example, the two most prominent topics on vaccine supporters side in period \emph{i} is mocking of the stringent anti-vaxxers with sarcastic posts including images.
Note that sarcasm often makes text-based opinion detection challenging, however we found no mis-classifications of such content during manual examination of the topics, proving the strength of the network-based approach.
As the time goes on, the supporters' topics progress from posts wishing for a vaccine, sharing news around its development, criticism around its roll-out (especially in the Lombardia region), and the final two periods are dominated by calls for people to get vaccinated.
The topical progression is much more varied on the hesitant side.
Period \emph{i} begins with the hashtag \texttt{\#lastragesilenziosa}, roughly translated as ``the silent slaughter'', with a link to a TV news story of a child dying shortly after being vaccinated.
Period \emph{ii}, instead, quotes an interview of an (unidentified) doctor who talks about elderly patients getting sick after a flu vaccine.
As the COVID-19 vaccine becomes prominent, a series of concerns dominate the hesitant side: microchips in vaccines (period \emph{iii}), ``experimental'' status of vaccines (\emph{iv}), and vaccine side effects and possible deaths (\emph{v-vi}).
We note that some of the concerns expressed by the hesitant community predate the actual enforcement by many months.
For instance, vaccination passports were being discussed by the community as early as February 2021, however the European agreed on a ``green pass'' only in April, and it wasn't until the summer of 2021 that first certificates have been issued\footnote{\url{https://euobserver.com/tickers/151543} \\ \url{https://www.gazzettaufficiale.it/eli/id/2021/06/17/21A03739/sg}}. 
Similarly, the calls against vaccinating children became prominent in period \emph{v}, whereas the vaccination of children of ages 5-11 did not take place in Italy until December 2021\footnote{\url{https://www.rainews.it/archivio-rainews/articoli/amp/ContentItem-e886e7fd-057d-4826-b5ce-7165b3a70dd2.html}} (after the end of the data collection).
Thus, we observe that the vaccine hesitant community may provide early indications of potential controversies around future public health decisions.

Finally, we examine the top cited URL domains by both sides.
In our analysis, we find the ranking does not change substantially across time periods, so we report the ranking computed over the whole duration.
The vaccination supporters mostly link to the large local newspapers, including (from top) Repubblica, Italian Huffington Post, TG COM24, and major news providers: Ansa and adnkronos.
At the top of the vaccination hesitant list we find \emph{imolaoggi.it}, a local news website with right-leaning content. 
The second most cited domain is YouTube, a resource widely known to be used by the anti-vaccination movement \cite{donzelli2018misinformation}.
The third domain is \emph{stopcensura.it}, espousing conspiratorial content, as well as linking deaths to COVID-19 vaccinations.
The rest is a mixture of domains as above and reputable news providers. 
Thus, although the top domains in both rankings are different between the two communities, there is an overlap in general news sources, providing a possible common ground. 
However, the context around these links must be further examined in order to reveal the interpretation by each side and selective sharing.







\section{Discussion \& Conclusions}

In this study, we show that the events of the COVID-19 pandemic only partially decrease the effects of the echo chambers around the vaccination debate on Twitter.
The division between who the vaccine supporters and hesitant users retweet, signifying attention and agreement, remains substantial throughout the first year and a half of the pandemic. 
Despite drastic changes in the focus around the topic of vaccination, we find that the membership of the two sides rarely mixes, with users continuing the same retweet behavior throughout the whole period of study.
The public health communication around the necessity and safety of the vaccines seems not to have allayed the concerns of vaccine hesitant, as the retweet community continued to grow into November 2021. 
These findings support previous observations that factual correction of information around vaccination does not change minds, and instead may decrease the intent to vaccinate~\cite{nyhan2015does}.
Furthermore, there are criticisms that unclear, mixed messaging from decision-makers and public health authorities has contributed to the skepticism around the unusual circumstances surrounding the vaccination campaign and other public health measures~\cite{turner2021poor}.
Examining the impact of official communication on the divisions of opinion in public discourse is an exciting future research direction.

Instead, the divisions in terms of mentions and topics of discussion have somewhat decreased during the vaccination campaign, potentially shifting the focus of the vaccination supporters and hesitant users to the same events and issues.
This is an encouraging development, as it is necessary to establish a commonly agreed-upon set of facts to foster the communication between the two sides.
Still, we see the hesitant community greatly increases in size over this time, adding thousands of users who have not previously engaged with the topic in our dataset (although they might have previously). 
A follow-up study of people engaging in the vaccine hesitant community in each subsequent time period may provide a profile of newly hesitant, and may help in understanding the concerns emerging due to the communication around COVID-19 vaccinations specifically. 
It is especially important not to assume that everybody engaging in the discussion is an ``anti-vaxxer'', and seeking opportunities for engagement and addressing the fears of those whose opinions are open to change, given correct intervention.

Beyond examining the dynamic nature of the vaccination debate, this paper contributes to the theoretical understanding of the echo chambers, considering the phenomenon from a network perspective, as well as from semantic focus of the posted content. 
By fitting topics modeled from one side to the other side, we show the convergence of topical focus of the two sides, beyond the retweeting or linking behaviors (as is focus in \cite{garimella2018political}). 
It is important, however, to monitor other discourse platforms, as it has been shown that those with different communication affordances, such as forum website Reddit \cite{morales2021no}, or specialized forum for parents \cite{betti2021detecting}, may not display as much polarization.

Moreover, we note a major limitation of this work -- its focus on the most active participants of the debate. 
The construction of the retweet networks favors the selection of those who have retweeted at least twice, discarding more than half of the users captured in the original collection. 
Indeed, it has been shown in previous literature that classifying the stance of the users who post few times (the ``silent majority'') is difficult \cite{cossard2020falling}, both manually and automatically, due to the lack of information on their position.
Still, the users captured in the retweet networks are responsible for 86\% of the tweets (``vocal minority''). 
Thus, the analysis shown here pertains more to those more engaged in the conversation, and should be used by policymakers in tandem with other methodologies, including surveys and interviews.


\subsection{Broader Impact \& Ethical Considerations}

Major efforts are ongoing to better understand the reasons for vaccine hesitancy in Europe and around the world. 
The European Centre for Disease Prevention and Control (ECDC) conducts regular surveys to measure COVID-19 behaviors, and supports the inclusion of additional data sources to understand the beliefs and expectations around vaccination \cite{ecdc2021facilitating}.
The surveillance tools and opinion community detection methods presented in this paper may assist in better understanding the popular sentiments, associations, questions, and misunderstandings that circulate on one of the most popular social media in Italy.
Although this dataset captures a small portion of Italian-speakers, most social media users do not engage in posting, and use platforms for entertainment and as a source of information \cite{van20141onepercent}, thus the potential audience of the content captured here may be an order of magnitude larger than the users we analyze.
Still, we should remember that there are other groups which are not captured in this data, including people not having access to internet and the platform, and those with disabilities not allowing them to interact with it.

While the dataset collected for this study contains only publicly posted tweets, it is possible that it contains posts from vulnerable groups, including those with serious or chronic health conditions relevant to COVID-19 and the vaccination campaigns, those emotionally or psychologically vulnerable, and family members and friends who are concerned for the wellbeing of their loved ones, among others. 
In order to preserve the privacy of the individual users, we reveal the Twitter handles only of public figures or anonymous accounts in this paper, and otherwise do not use identifiable information in the analysis of the data. 
Furthermore, we will abide by Twitter's Terms of Service\footnote{\url{https://developer.twitter.com/en/developer-terms/}} by making the IDs of the collected available upon request, such that those which have been deleted by their authors will not be available when the metadata is re-collected (unfortunately, limiting reproducibility somewhat).
However, to support the transparency of this work, we make the code available to the research community, especially that pertaining community detection, random walk controversy score computation, and community-aware topic modeling\footnote{\url{https://github.com/ISIFoundation/Vax-IT-20-21}}.
Finally, as opinion surveillance methods used in this work may be applied to identify other communities and individuals therein, we urge the research community to strictly abide by the code of ethics for the research and application of these tools (such as one by AAAI\footnote{\url{https://www.aaai.org/Conferences/code-of-ethics-and-conduct.php}}) in order to minimize harm to the subjects of research.

\section{Acknowledgments}

The authors acknowledge support from the Lagrange Project
of the Institute for Scientific Interchange Foundation (ISI
Foundation) funded by Fondazione Cassa di Risparmio di
Torino (Fondazione CRT).
AP acknowledges partial support from Intesa Sanpaolo Innovation Center. The funders had no role in study design, data collection and analysis, decision to publish, or preparation of the manuscript. This work has also received funding from the European Union's Horizon 2020 research and innovation programme - project EpiPose (Grant agreement number 101003688) and project PANDEM-2 (Grant agreement number  883285). This work reflects only the authors’ view. The European Commission is not responsible for any use that may be made of the information it contains.

\bibliography{bibliography}

\end{document}